\documentclass[10pt,conference]{IEEEtran}                                                           \IEEEoverridecommandlockouts

\usepackage{xspace,empheq,fancybox,amsmath,amssymb,epsfig,subfigure,amsthm}
\usepackage{psfrag,color,bm}
\usepackage{url,cite,footnote,xspace,algorithm,algorithmic}
\usepackage{verbatim}
\usepackage[T1]{fontenc}

\allowdisplaybreaks[3]
\input{mysymbol.sty}

\newcommand{\ind}{1\hspace{-1.3mm}1}
\newcommand{\rank}{{\textrm{rank}}}

\begin{document}

\title{Robust Power System State Estimation\\
        for the Nonlinear AC Flow Model}

\author{\large Hao Zhu,~\IEEEmembership{Student Member, IEEE},
and Georgios B. Giannakis,~\IEEEmembership{Fellow, IEEE}
\thanks{\protect\rule{0pt}{1em}
The authors are with the Dept. of ECE and the Digital Technology
Center, University of Minnesota, Minneapolis, MN 55455, USA.
Tel/fax: (612)624-9510/625-2002, emails:
{\tt \{zhuh,georgios\}{\rm\char64}umn.edu}}%
}

\maketitle

%

\begin{abstract}
\noindent An important monitoring task for power systems is accurate
estimation of the system operation state. Under the nonlinear AC
power flow model, the state estimation (SE) problem is inherently
nonconvex giving rise to many local optima. In addition to
nonconvexity, SE is challenged by data integrity and cyber-security
issues. Unfortunately, existing robust (R-) SE schemes employed
routinely in practice rely on iterative solvers, which are sensitive
to initialization and cannot ensure global optimality. A novel R-SE
approach is formulated here by capitalizing on the sparsity of an
overcomplete outlier vector model. Observability and identifiability
issues of this model are investigated, and neat links are
established between R-SE and error control coding. The \emph{convex}
semidefinite relaxation (SDR) technique is further pursued to render
the nonconvex R-SE problem efficiently solvable. The resultant
algorithm markedly outperforms existing iterative alternatives, as
corroborated through numerical tests on the standard IEEE 30-bus
system.
\end{abstract}


\section{INTRODUCTION}
The electric power grid is a complex cyber-physical system
consisting of multiple modules, each with a transmission
infrastructure spanning over a huge geographical area, transporting
energy from generation sites to distribution networks. Monitoring
the operational conditions of grid transmission networks is of
paramount importance to facilitate system control and optimization
tasks, including security analysis and economic dispatch with
security constraints; see e.g., \cite[Ch. 1]{SE_book} and
\cite{monieee00}. For this purpose, various system variables are
measured in distant buses and then transmitted to the control center
for estimating the system state variables, namely complex bus
voltages. Due to the wide spread of transmission networks and the
current integration of enhanced computer/communication
infrastructure, the power system state estimation (SE) is challenged
by data integrity concerns arising due to ``anomalous'' measurements
affected by \emph{outliers} \cite{xu_arxiv11,vkgg_tps12} and/or
adversarial \emph{cyber-attacks} \cite{bobba_ccs09,kosut_tsg11,liu_ccs09}. These
concerns motivate the development of robust approaches to improve
resilience of SE to anomalous (a.k.a. bad) data.

For the AC power flow model however, SE challenges come not only
from anomalous data, but are further magnified due to the nonlinear
couplings present between meter measurements and state variables. To
cope with these challenges, Gauss-Newton iterative solvers estimate
the state of an approximate linear regression model per iteration,
using robust renditions of the weighted least-squares (WLS) error
criterion, such as the weighted least-absolute value (WLAV) one; see
e.g., \cite[Ch. 2-6]{SE_book}. The current iteration adopts Taylor's
first-order expansion around the estimate of the previous iteration
to approximate the quadratic AC model with the aforementioned linear
regression model. This iterative procedure is closely related to
gradient descent algorithms for solving nonconvex problems, see
e.g., \cite[Ch. 1]{Bertsekas-NonlinProg}, which are known to
encounter two issues: i) sensitivity to initialization; and ii)
convergence concerns. Existing variants have asserted improved
numerical stability of the iterative procedure~\cite[Ch.
3]{SE_book}. Latest SE trends incorporate \emph{linear} state
measurements offered by synchronized phasor measurement units
(PMUs); see e.g., \cite{gomezpscc11} and references therein.
However, limited PMU deployment currently confines SE to mostly rely
on the traditional nonlinear meter measurements, and its companion
Gauss-Newton iterative methods. Hence, it is very important to
develop a robust (R-) SE solver tailored for the nonlinear
measurement model, and capable of approximating the \textit{global
optimum} at \textit{polynomial complexity}.


The present paper adopts an overcomplete additive outlier-aware
measurement model, and leverages the sparsity of outliers to develop
an R-SE approach to jointly estimate system states and identify the
outliers present (Section \ref{sec:ps}).

Inherent to the overcomplete outlier-aware model is system
under-determinacy, which in turn raises outlier observability and
identifiability concerns (Section \ref{sec:id}). It was recently
recognized that there are unobservable cyber-attacks that the system
operator would fail to detect \cite{bobba_ccs09,liu_ccs09,kosut_tsg11}, but all
studies so far are limited to linear approximate SE models.
Theoretical guarantees of the sparse outlier model were explored for
nonlinear SE models in \cite{xu_arxiv11}. Compared to these works,
the present one provides a unifying framework to understand how
tolerant the nonlinear regression model is to data corruption, by
introducing the notion of \emph{measurement distance}. The latter is
nicely connected to distance metrics popular in channel coding
theory, which are known to determine the error-control capability of
channel codes. This connection reveals why the measurement distance
is instrumental to characterizing the regression function's
resilience to outliers.

In addition, the novel R-SE framework lends itself to a convex
relaxation approach, which yields R-SE solvers approximating the
global optimum (Section \ref{sec:sdr}). A well appreciated tool for
convexifying non-convex problems \cite{luospmag10}, semidefinite
relaxation (SDR) solvers thus emerge as powerful schemes for R-SE of
nonlinear AC power flow models. Preliminary tests on the IEEE 30-bus
system corroborate the performance improvement of the proposed
approach (Section \ref{sec:sim}).

\vspace{1mm} \noindent {\it Notation:} Upper (lower) boldface
letters will be used for matrices (column vectors); $(\cdot)^T$
denotes transposition; $(\cdot)^\ccalH$ complex-conjugate
transposition; Re$(\cdot)$ the real part; Im$(\cdot)$ the imaginary
part; Tr$(\cdot)$ the matrix trace; rank$(\cdot)$ the matrix rank;
$\mathbf 0$ the all-zero matrix; $\|\cdot\|_p$ the vector $p-$norm
for $p\geq 1$; $\lfloor \cdot \rfloor$ the floor of a real number; and $|\cdot|$ ($\measuredangle$) the magnitude
(angle) of a complex number.


\section{Modeling and Problem Statement}
\label{sec:ps}

Consider a power transmission network with $N$ buses denoted by the
set of nodes $\ccalN := \{1,\ldots,N\}$, and $L$ transmission lines
represented by the set of edges $\ccalE := \{(n,m)\} \subseteq
\ccalN \times \ccalN$. Suppose $M$ measurements are taken for
estimating the complex voltage states $\{V_n\}_{n\in \ccalN}$, from
a subset of the following system variables:


\begin{itemize}

\item $P_{n}(Q_{n})$: the real (reactive) power injection at bus $n$ (negative if bus $n$ is connected to a load);

\item $P_{mn}(Q_{mn})$: the real (reactive) power flow from bus $m$ to bus $n$; and

\item $|V_n|$: the voltage magnitude at bus $n$.

\end{itemize}

Compliant with the AC power flow model \cite{PSA_book}, these
measurements obey nonlinear equations relating them with the system
state vector $\bbv := [V_1,\ldots,V_N]^T \in \mathbb C^N$. These
equations also involve the injected currents of all buses that are
here collected in the vector $\bbi:=[I_1,\ldots,I_N]^T \in \mathbb
C^N$, as well as the currents, flowing from say bus $m$ to $n$,
denoted by $I_{mn}$. Kirchoff's law in vector-matrix form simply
dictates $\bbi = \bbY \bbv$, where $\bbY \in \mathbb C^{N\times N}$
denotes the grid's symmetric bus admittance matrix having $(m,n)$-th
entry given by
\begin{align}\label{ymn}
Y_{mn} :=
\left \{\begin{array}{ll} -y_{mn}, &\textrm{if }(m,n)\in \ccalE
\\y_{nn} + \sum_{\nu \in \ccalN_n} y_{n\nu}, & \textrm{if } m=n
\\0, & \textrm{otherwise} \end{array} \right.
\end{align}
with $y_{mn}$ denoting the line admittance between buses $m$ and
$n$; $y_{nn}$ bus $n$'s admittance to the ground; and $\ccalN_n$ the
set of all buses linked to bus $n$ through transmission lines. In
addition, the current flow is given by $I_{mn} = \bbary_{mn} V_m +
y_{mn} (V_m-V_n)$, with $\bbary_{mn}$ standing for the shunt
admittance at bus $m$ associated with line $(m,n)$. Clearly, all
current variables are linearly related to the state $\bbv$. As for
the nonlinear measurements, the AC power flow model asserts that the
apparent power injection into bus $n$ is given by $P_n + jQ_n = V_n
I_n^\ccalH$, while the apparent power flow from bus $m$ to bus $n$
by $P_{mn}+ jQ_{mn} = V_m I_{mn}^\ccalH$. Further, expressing the
squared bus voltage magnitude as $|V_n|^2 = V_n V_n^\ccalH$, it is
clear that all measurable quantities listed earlier are nonlinearly
(in fact quadratically) related to $\bbv$.

Apart from the nonlinearity present, another challenge present in
the SE is due to grossly corrupted meter measurements (a.k.a. bad
data). Statistical tests such as the largest normalized residuals of
the weighted least-squares (WLS) estimation error are typically
employed to reveal and remove bad data \cite{monieee00}.
Alternatively, robust estimators, such as the least-absolute
deviation, or Huber's M-estimators have also been considered; see
e.g., \cite[Ch. 6]{SE_book}. Motivated by recent advances in
sparsity-aware robust statistical inference \cite{uspacor}, the
fresh look at robust SE (R-SE) advocated in this paper is by an
overcomplete model for the outlying data. To this end, collect first
the $M$ measurements in the vector $\bbz :=$ $[\{\check
P_{n}\}_{n\in\ccalN_P},$ $\{\check Q_n\}_{n\in\ccalN_Q},$ $\{\check
P_{mn}\}_{(m,n) \in \ccalE_{P}},$ $\{\check Q_{mn}\}_{(m,n) \in
\ccalE_{Q}},$ $\{|\check V_n|^2\}_{n\in\ccalN_V}]^T$, where the
check mark differentiates measured values from the noise-free
variables\footnote{For consistency with other measurements,
$|V_n|^2$ is considered from now on. This is possible by adopting
$|\check V_n| = |V_n| +\epsilon_V$, where $\epsilon_V$ is zero-mean
Gaussian with small variance $\sigma_V^2$, to obtain the approximate
model $|\check V_n|^2 \approx |V_n|^2 +\epsilon_V'$, where
$\epsilon_V'$ has variance $4|\check V_n|^2 \sigma_V^2$.}. Consider
also the scalar variables $\{a_\ell\}_{\ell=1}^M$ one per
measurement, taking the value $a_\ell = 0$ if the $\ell$-th
measurement obeys the nominal (outlier-free) model, and $a_\ell \neq
0$ if it corresponds to a bad datum. This way, the nonlinear
measurement model becomes
\begin{align}
z_\ell = h_\ell(\bbv) + \epsilon_\ell + a_\ell, ~~\ell=1,\ldots, M \label{zell}
\end{align}
where $h_\ell(\cdot)$ captures the quadratic relationship specified
by the aforementioned AC power flow equations, and the zero-mean
additive Gaussian white noise (AWGN) $\epsilon_\ell$ is assumed
uncorrelated across meters with variance $\sigma_\ell^2$.

Recovering both $\bbv$ and the $M\times 1$ vector $\bba:=[a_1,
\ldots, a_M]^T$ essentially reveals the state and identifies faulty
measurements. However, the system in \eqref{zell} with both $\bbv$
and $\bba$ being unknown is under-determined, as the number of
measurements $M$ is always less than the number  of unknowns $N+M$.
Instrumental to handling this under-determinacy will be the
(arguably low) percentage of outliers, which gives rise to a (high)
\emph{level of sparsity}, that is the number of zero entries in
$\bba$. The degree of sparsity will be further linked in the ensuing
section with the notions of observability and identifiability of the
outlier vector. By capitalizing on the sparsity of $\bba$, the goal
of jointly estimating and identifying $\bbv$ and $\bba$ can be
achieved by the following outlier-sparsity-controlling criterion
%
\begin{align}
\{\hhatbbv,\hhatbba\} :=& \arg \min_{\bbv,\bba} \sum_{\ell=1}^M
w_\ell ~[z_\ell - h_\ell (\bbv) - a_\ell]^2 + \lambda \|\bba\|_0
\label{rSE_wls0}
\end{align}
where $w_\ell := 1/\sigma_\ell^2$ $\forall \ell$, and $\lambda >0$
scales the regularization term which comprises the
$\ell_0$-pseudonorm, i.e., the number of non-zero $a_\ell$'s that
naturally controls the number of outliers in $\hhatbba$. Even with
linear models however, solving the optimization problem in
\eqref{rSE_wls0} is NP-hard due to the $\ell_0$-norm
regularization~\cite{ct06tit}. Before proposing efficient schemes
for solving the under-determined problem in \eqref{rSE_wls0}, the
next section will provide observability and identifiability analysis
to assess the ability of R-SE to cope with sparse outlier patterns.

\section{Outlier Observability and Identifiability}
\label{sec:id}

The goal of this section is to investigate fundamental uniqueness
issues associated with the system under-determinacy arising due to
the overcomplete outlier-aware model in \eqref{zell}. To isolate
uniqueness from noise resilience issues, focus is placed on the
\emph{noise-free} outlier-aware measurement model written in vector
form as
\begin{align}
\bbz = \bbh(\bbv) + \bba \label{vecz}
\end{align}
with the high-dimensional function $\bbh(\cdot): \mathbb C^N
\rightarrow \mathbb R^M$.

\begin{definition}\label{def:id}
Given measurements $\bbz = \bbh(\bbv_o) + \bba_o$, with $\bbv_o$
denoting the true state, and $\bbh(\cdot)$ known, the outlier vector
$\bba_o$ is \textit{observable}  if and only if (iff)
$\forall~\bbv_o$ the set
\begin{align}
\ccalV:= \{\bbv\in \mathbb C^N |~ \bbz= \bbh(\bbv_o) + \bba_o = \bbh(\bbv)\}  \label{setv}
\end{align}
is empty. Furthermore, the outlier vector $\bba_o$ is
\textit{identifiable} iff $\forall~\bbv_o$ the set
\begin{align}
\ccalS : = \{(\bbv,\bba)|~ \bbh(\bbv)+\bba = \bbz, ~ \|\bba\|_0\leq \|\bba_o\|_0 \} \label{sets}
\end{align}
has only one element, namely $(\bbv_o,\bba_o)$.
\end{definition}

But why are outlier observability and identifiability intuitively
important? For an observable $\bba_o$, upon collecting $\bbz$, the
system operator can discern whether there are bad data or not. In
addition, for an identifiable $\bba_o$, the system operator can
recover exactly (in the absence of nominal noise) \emph{both}
$\bba_o$ and $\bbv_o$ in the presence of bad data.

Definition \ref{def:id} implies that if $\bba_o$ is identifiable,
then it is necessarily observable, because otherwise the set
$\ccalV$ in \eqref{setv} would have at least one element $\bbv' \in
\mathbb C^N$; in which case, the pair $(\bbv', \bb0)$ would be an
additional second element of $\ccalS$ in \eqref{sets} - a fact
contradicting identifiability. Therefore, as a property of an
outlier vector $\bba_o$ identifiability is stronger than (i.e.,
subsumes) its observability.

Without accounting for the nominal AWGN in \eqref{zell}, it is
possible to reduce the cost in \eqref{rSE_wls0} to only the
$\ell_0$-norm, while including the quadratic part as  equality
constraint to obtain
\begin{align}
\{\hhatbbv,\hhatbba\} :=& \arg \min_{\bbh(\bbv)+\bba = \bbz}  \|\bba\|_0  \nonumber\\
 =& \arg\min_{\bbv,\bba= \bbz-\bbh(\bbv)} \|\bbz-\bbh(\bbv)\|_0.\label{rSE_noise}
\end{align}
Clearly, for the noise-free R-SE problem in \eqref{rSE_noise}, the
pair $(\bbv_o, \bba_o)$ is feasible, and the cost evaluated at
$(\bbv_o, \bba_o)$ equals $\|\bbz-\bbh(\bbv_o)\|_0 = \|\bba_o\|_0$.
This is also the minimum cost attainable
when $\bba_o$ is identifiable, as there is no other pair $(\bbv,
\bba)$ with smaller $\ell_0$-norm $\|\bba\|_0$ according to
Definition \ref{def:id}. Conversely, if the noise-free problem
\eqref{rSE_noise} has a unique solution given by $(\bbv_o, \bba_o)$,
then $\bba_o$ is identifiable. Similarly, the noise-free R-SE
formulation can easily detect the presence of bad data if the
minimum achievable is non-zero. This clearly demonstrates the role
of the outlier vector's $\ell_0$-norm  in the R-SE criterion
\eqref{rSE_wls0}, in identifying the presence of bad data, or, in
recovering the true state even when bad data are present.

A critical attribute for an observable (identifiable) outlier vector
is its maximum sparsity level $K_o$ (respectively $K_i$). To
appreciate this, consider the two broad classes that outliers typically come from. The first class includes bad data emerging due to faulty
meters, telemetry errors, or software bugs, which generally occur
rarely, that is with low probability; see e.g., \cite{uspacor} and
references therein. Here, $K_o$ quantifies the maximum number of bad
data that can be revealed with high probability; while $K_i$ denotes
the maximum number of outlying meters that can be identified so that
recovery of the true state becomes feasible. The second source of
outliers comprises malicious data attacks, in which the adversary
can typically control only a subset of meters with limited
cardinality \cite{bobba_ccs09,liu_ccs09,kosut_tsg11}. In this class of outliers,
$K_o$ and $K_i$ can suggest the minimum number of meters that must
be protected to render malicious data attacks ineffective.

Even though $K_o$ ($K_i$) is useful for assessing the degree of
outlier observability (identifiability), deciding whether a given
vector $\bba_o$ is observable or identifiable for the nonlinear AC
model \eqref{vecz} is challenging, except for the trivial case
$\bba_o = \bb0$. Fortunately, it is possible to obtain $K_o$ and
$K_i$ by leveraging the notion of the measurement distance for any
nonlinear function $\bbh(\cdot)$, as defined next.

\begin{definition}\label{def:dist}
The \textit{measurement distance} for the function $\bbh(\cdot):
\mathbb C^N \rightarrow \mathbb R^M$ is given by
\begin{align}
D(\bbh) &:= \min_{\bbv\neq\bbv'} \|\bbh(\bbv) -\bbh(\bbv')\|_0  \nonumber\\
&= \min_{\bbv\neq\bbv'} \sum_{\ell=1}^M \ind[h_\ell(\bbv)-h_\ell(\bbv')] \label{dist_def}
\end{align}
where $\ind$ denotes the indicator function.
\end{definition}

The notion of measurement distance parallels that of the Hamming
distance in channel coding theory; see e.g., \cite[Sec.
7.11]{cover_eit06}. Given any linear mapping over a known finite
field, the Hamming distance characterizes the minimum difference
between any two strings that lie in the mapped space, and it can be
easily computed for fixed problem dimensions. However, for the R-SE
problem of interest, $\bbv$ is drawn from the complex field $\mathbb
C^N$, while the mapping $\bbh(\cdot)$ is quadratic. Compared to the
Hamming distance it will be generally very challenging to compute
$D(\bbh)$ in \eqref{dist_def}.

Interestingly, as the Hamming distance has been popular due to its
connection with the error control capability of linear channel
codes, the measurement distance in \eqref{dist_def} will turn out to
be particularly handy in characterizing outlier observability and
identifiability, as asserted in the following proposition.

\begin{proposition}\label{prop:K}
Given the measurement distance $D$ for the nonlinear function
$\bbh(\cdot)$ in \eqref{dist_def}, the maximum sparsity level of an
observable outlier vector is $K_o = D-1$, while the maximum one of
an identifiable outlier vector is $K_i = \lfloor \frac{D-1}{2}
\rfloor$.
\end{proposition}

The proof for both statements follows readily from Definition
\ref{def:id} using simple contradiction arguments, and for this
reason it is omitted. Notice that the second part can also be
deduced after adapting \cite[Thm. 5.1]{xu_arxiv11}, which neither
explicitly relates to the notion of measurement distance, nor it is
linked with the maximum sparsity level of observable outliers.

Using the measurement distance metric, Proposition \ref{prop:K}
provides a unifying framework to understand the tolerance of any
function $\bbh(\cdot)$ to the number of outlying data. Since the
measurement distance of any nonlinear function is difficult to
obtain, the ensuing subsection pursues linearized approximants of
the quadratic measurement model, which are typically employed by
Gauss-Newton iterative SE solvers, and can be used to provide
surrogate distance metrics. Depending on initialization, the linear
approximants could not only be very accurate, but will also shed
light on understanding uniqueness issues associated with nonlinear
AC power system models.

\subsection{Linear Approximation Model}
\label{sec:linear}

Consider linearizing the nonlinear measurement model \eqref{vecz}
expressed in terms of the polar coordinates of the state vector, as
in e.g., \cite[Sec. 2.6]{SE_book}. Toward this end, the $N\times 1$
complex vector $\bbv$ is mapped first to the $2N\times 1$ real
vector $\bbx : = [|V_1|,\ldots, |V_N|, \measuredangle V_1, \ldots,
\measuredangle V_N] \in \mathbb R^{2N}$. Invoking the first-order
Taylor expansion, the noise-free $\bbz$ can be approximated around a
given point $\bbarbbv$, or the corresponding $\bbarbbx$, by
\begin{align}
\bbz = \bbh(\bbv) +\bba \approx  \bbh(\bbarbbv) +
\bbH_\bbarx (\bbx-\bbarbbx)  +\bba \label{zapprox}
\end{align}
where $\bbH_\bbarx \in \mathbb R^{M\times (2N)}$ denotes the
Jacobian matrix evaluated at $\bbarbbx$. Upon defining $\tdbbz : =
\bbz-\bbh(\bbarbbv) + \bbH_\bbarx \bbarbbx$,  the approximate model
\eqref{zapprox} becomes a linear one in the unknown $\bbx$, that is
\begin{align}
\tdbbz  \approx  \bbH_\bbarx \bbx  +\bba.  \label{zlinear}
\end{align}

The measurement distance of the linear function in \eqref{zlinear}
can be found easily, as summarized next.

\begin{proposition}\label{prop:lin}
The measurement distance for any linear mapping characterized by a full column-rank matrix
$\bbH_\bbarx \in \mathbb R^{M\times (2N)}$ is $D = M+1-${\rm
\rank}$(\bbH_\bbarx)$.
\end{proposition}

The proof relies on simple linear algebra arguments as follows.
Using Definition \ref{def:dist}, the measurement distance $D:=
\min_{\bbx-\bbx'\neq \bb0} \|\bbH_\bbarx(\bbx-\bbx')\|_0$ is
attained when matrix $\bbH_\bbarx$ has at most $(M-D)$ linearly
dependent rows; otherwise, the number of zero entries of
$\bbH_\bbarx(\bbx-\bbx')$ would be $(M-D+1)$ and that of non-zero ones $(D-1)$, which leads to a contradiction; hence, rank($\bbH_\bbarx$)$=M-D+1$, as asserted by
Proposition \ref{prop:lin}.

Recalling from Proposition \ref{prop:K} how $D$ is linked with the
outlier observability and identifiability levels, the next corollary
follows readily.

\begin{corollary}\label{cor:lin}
For any linear mapping characterized by $\bbH_\bbarx$, the maximum
sparsity level of an observable outlier is $K_o = M-${\rm
\rank}$(\bbH_\bbarx)$, while the maximum sparsity level of an
identifiable outlier is $K_i = \lfloor
\frac{M-\mathrm{rank}(\bbH_\bbarx)}{2} \rfloor$.
\end{corollary}

For the linear approximation model in \eqref{zlinear}, the
measurement distance $D$ grows linearly with the  number of meters
$M$. This demonstrates that measurement redundancy is very
beneficial for improving resilience to outliers. Conceivably, $D$
could be further boosted thanks to the nonlinearity in
$\bbh(\cdot)$. Compared to its linear counterpart, the quadratic
function $\bbh(\cdot)$ is likely to increase the dimension of the
space that is mapped to, and thus lead to a larger measurement
distance in a space of higher dimensionality. This is precisely the
reason why highly nonlinear functions find important applications to
cryptography \cite{carlet_jc04}.
Although linearization provides a viable approximant, quantifying
(or bounding) the measurement distance for the quadratic
$\bbh(\cdot)$ corresponding to the AC power flow model constitutes
an interesting future research direction.

\section{Solving the R-SE via SDR}
\label{sec:sdr}

This section will leverage convex relaxation techniques to solve the
R-SE problem in \eqref{rSE_wls0}. First, building on the premise of
compressive sampling \cite{ct06tit}, the $\ell_1$-norm can be
employed to tackle the NP-hard $\ell_0$-norm and relax the R-SE cost
in \eqref{rSE_wls0} to
%
\begin{align}
\{\hhatbbv,\hhatbba\} :=& \arg \min_{\bbv,\bba} \sum_{\ell=1}^M w_\ell
~[z_\ell - h_\ell (\bbv) - a_\ell]^2 + \lambda \|\bba\|_1. \label{rSE_wls}
\end{align}
The $\ell_1$-norm relaxation has been used for various robust
statistical inference tasks, with documented theoretical guarantees;
see e.g., \cite{uspacor}. It has also been adopted recently for R-SE
with the linear measurement model in \cite{vkgg_tps12}, and also
with the nonlinear one in \cite{xu_arxiv11} using the linearization
technique employed by iterative Gauss-Newton SE solvers. However,
the pertinent performance analysis in \cite{xu_arxiv11} has been
given for a general nonlinear regression model, but not for the
specific quadratic measurement model corresponding to AC power
systems. Moreover, it is worth stressing that the iterative
optimization framework adapted in \cite{xu_arxiv11} to solve
\eqref{rSE_wls} offers no guarantees regarding convergence or global
optimality. In a nutshell, the desiderata remains to develop an R-SE
solver capable of accounting for the practical AC quadratic
measurement model, while attaining or approximating the
\textit{global optimum} at \textit{polynomial-time} complexity.

This task will be pursued here using semidefinite relaxation (SDR),
which has been recently recognized as a powerful technique for
convexifying the SE with nonlinear measurement models \cite{hzgg_naps11}.
To this end, each quadratic measurement $z_\ell$ will be expressed
linearly in terms of the outer-product matrix 
$\bbV:=\bbv\bbv^\ccalH$.
Let $\{\bbe_n\}_{n=1}^N$ denote the canonical basis of $\mathbb
R^{N}$, and define the following admittance-related matrices
\begin{subequations}
\label{Ymats}
\begin{align}
\bbY_n &:= \bbe_n \bbe_n^T\bbY \label{Yn}\\
\bbY_{mn} &:= (\bbary_{mn} + y_{mn}) \bbe_m \bbe_m^T - y_{mn}
\bbe_m\bbe_n^T  \: \label{Ymn}
\end{align}
\end{subequations}
and their related Hermitian counterparts
\begin{subequations}
\label{Hmats}
\begin{align}
\bbH_{P,n} &\!:=\! \frac{1}{2}\left(\bbY_n+\bbY_n^\ccalH\right), ~~~ \bbH_{Q,n} \!:=\! \frac{j}{2}\left(\bbY_n-\bbY_n^\ccalH\right)  \label{PQn}\\
\bbH_{P,mn} &\!:=\! \frac{1}{2}\!\left(\!\bbY_{mn}\!+\!\bbY_{mn}^\ccalH\!\right), 
\bbH_{Q,mn} \!:=\! \frac{j}{2}\left(\!\bbY_{mn}\!-\!\bbY_{mn}^\ccalH \!\right) \label{PQmn}\\
\bbH_{V,n} &:= \bbe_n \bbe_n^T. \label{Vn}
\end{align}
\end{subequations}
Using these definitions, the following lemma is proved in
\cite{hzgg_naps11} to establish a linear model in the complex
rank-one matrix $\bbV$.

\begin{lemma} \label{lem:linX} All error-free measurement
variables are linearly related with the outer-product $\bbV$ as
\begin{subequations}
\label{trX}
\begin{align}
P_n &= \mathrm{Tr}(\bbH_{P,n} \bbV), ~~~~~
Q_n = \mathrm{Tr}(\bbH_{Q,n} \bbV) \label{trXPQn}\\
P_{mn} &= \mathrm{Tr}(\bbH_{P,mn} \bbV),~
Q_{mn} = \mathrm{Tr}(\bbH_{Q,mn} \bbV) \label{trXPQmn}\\
|V_n|^2 &= \mathrm{Tr}(\bbH_{V,n} \bbV). \label{trXVn}
\end{align}
\end{subequations}
Thus, the measurement $z_\ell$ in \eqref{zell} can be written as
\begin{align}
z_\ell = h_\ell(\bbv)  +\epsilon_\ell +a_\ell =\mathrm{Tr}(\bbH_\ell \bbV) + \epsilon_\ell +a_\ell \label{zltrX}
\end{align}
where $\bbH_\ell$ is a Hermitian matrix specified in accordance with
\eqref{PQn}-\eqref{Vn}.
\end{lemma}

Lemma \ref{lem:linX} implies the following equivalent reformulation
of \eqref{rSE_wls} [cf. \eqref{zltrX}]
\begin{subequations}
\label{SE_wls2}
\begin{align}
 \left\{\hhatbbV_1,\hhatbba_1\right\} &:= \arg\min_{\bbV,\bba}  \sum_{\ell=1}^M w_\ell \left[z_\ell - \textrm{Tr}(\bbH_\ell\bbV) -a_\ell\right]^2 +\lambda \|\bba\|_1 \label{SE_wls2f}\\
\textrm{s.to}& ~ \bbV \in \mathbb C^{N\times N} \succeq \bb0, \textrm{and rank}(\bbV) = 1 \label{SE_wls2c}
\end{align}
\end{subequations}
where the positive semi-definiteness and rank constraints jointly
ensure that for any $\bbV$ admissible to \eqref{SE_wls2c}, there
always exists a state vector $\bbv \in \mathbb C^N$ such that $\bbV
= \bbv\bbv^\ccalH$.

Albeit the linearity between $z_\ell$ and $\bbV$ in the new
formulation \eqref{SE_wls2}, nonconvexity is still present in two
aspects: i) the cost in \eqref{SE_wls2f} has degree 4 wrt the
entries of $\bbV$; and ii) the rank constraint in \eqref{SE_wls2c}
is nonconvex. Aiming for a semidefinite programming (SDP)
formulation of \eqref{SE_wls2}, Schur's complement lemma, see e.g.,
\cite[Appx. 5.5]{Boyd-Convex}, can be leveraged to convert the
summands in \eqref{SE_wls2f} to a linear cost over an auxiliary
vector $\bbchi \in \mathbb R^M$. Specifically, with
$\bbw:=[w_1,\ldots,w_L]^T$ and likewise for $\bbchi$, consider an
R-SE reformulation as
\begin{subequations}
\label{SE_sdpo}
\begin{align}
\left\{\hhatbbV_2,\hhatbba_2,\hhatbbchi_2\right\}:= 
&\arg \min_{\bbV,\bba,\bbchi} \bbw^T\bbchi +\lambda \|\bba\|_1 \hfill~ \label{SE_sdpof}\\
\textrm{s.to} ~~~  \bbV \succeq \bb0, & \textrm{and rank}(\bbV) = 1,  \label{SE_sdpoc}\\
&\hspace{-20mm}\left[\begin{array}{cc} -\chi_\ell & z_\ell- \textrm{Tr}(\bbH_{\ell} \bbV)-a_\ell \\ z_\ell- \textrm{Tr}(\bbH_{\ell} \bbV)-a_\ell & -1 \end{array} \right]  \preceq \bb0~\forall \ell. \label{SE_sdpocc}
\end{align}
\end{subequations}
Upon adapting results from \cite{hzgg_naps11}, the equivalence among
all three R-SE formulations can be asserted as follows.

\begin{proposition} \label{prop:SE}
For the AC power flow model, all three nonconvex formulations in
\eqref{rSE_wls}, \eqref{SE_wls2}, and \eqref{SE_sdpo}, solve an
equivalent R-SE problem. For the optima of these problems, it holds
that
\begin{align}\label{SE_eqv}
\hspace*{-0.2cm}
\hhatbbV_1 &= \hhatbbV_2= \hhatbbv\hhatbbv^\ccalH 
~~\mathrm{and}~ \hat \chi_{2,\ell} = \left[\hhatz_\ell -
\mathrm{Tr}(\bbH_\ell\hhatbbV_2)\right]^2~\forall \ell.
\end{align}
\end{proposition}

Proposition \ref{prop:SE} establishes the relevance of the novel
R-SE formulation \eqref{SE_sdpo}, which is still nonconvex though,
due to the rank-1 constraint. Fortunately though \eqref{SE_sdpo} is
amenable to the SDR technique, which amounts to dropping the rank
constraint and has well-appreciated merits as an optimization tool;
see e.g., \cite{luospmag10} for a tutorial treatment of its
applications in signal processing and communications. The
contribution here consists in permeating the benefits of this
powerful optimization tool to estimating the state of AC power
systems, even when outliers (bad data or cyber-attacks) are present.

In the spirit of SDR, relaxing the rank constraint in
\eqref{SE_sdpoc} leads to the following SDP formulation:
\begin{subequations}
\label{SE_sdp}
\begin{align}
&\left\{\hhatbbV,\hhatbba, \hhatbbchi\right\}:=\arg \min_{\bbV,
\bba,\bbchi} \bbw^T\bbchi +\lambda\|\bba\|_1 \label{SE_sdpf}\\
& ~~\textrm{s.to} ~ \bbV \succeq \bb0,  \label{SE_sdpc}\\
& \hspace{3mm}\left[\begin{array}{cc} -\chi_\ell & z_\ell-
\textrm{Tr}(\bbH_{\ell} \bbV)-a_\ell \\ z_\ell- \textrm{Tr}(\bbH_{\ell}
\bbV)-a_\ell & -1 \end{array} \right]  \preceq \bb0~\forall \ell.
\label{SE_sdpcc}
\end{align}
\end{subequations}

SDR endows R-SE with a convex SDP formulation for which efficient
schemes are available to obtain the global optimum using, e.g., the
interior-point solver SeDuMi \cite{sedumi}. The worst-case
complexity of this SDP solver is $\ccalO(M^4 \sqrt{N}
\log(1/\epsilon))$ for a given solution accuracy $\epsilon>0$
\cite{luospmag10}.  For typical power networks, $M$ is in the order
of $N$, and thus the worst-case complexity becomes
$\ccalO(N^{4.5}\log(1/\epsilon))$. Further computational complexity
reduction is possible by exploiting the sparsity, and the so-called
``chordal'' data structure of matrix $\bbV$, as detailed in
\cite{hzgg_naps11}.

Nonetheless, the SDP problem \eqref{SE_sdp} is only a relaxed
version of the equivalent R-SE in \eqref{SE_sdpo}; hence, its
solution $\hhatbbV$ may have rank greater than 1, which makes it
necessary to recover a feasible estimate $\hhatbbv$ from $\hhatbbV$.
This is possible by eigen-decomposing $\hhatbbV = \sum_{i=1}^{r}
\lambda_i \bbu_i\bbu_i^\ccalH$, where $r :=$ rank$(\hhatbbV)$,
$\lambda_1 \geq \cdots \geq \lambda_r >0$ denote the positive
ordered eigenvalues, and $\{\bbu_i \in \mathbb C^N\}_{i=1}^r$ are
the corresponding eigenvectors. Since the best (in the minimum-norm
sense) rank-one approximation of $\hhatbbV$ is $\lambda_1
\bbu_1\bbu_1^\ccalH$, the state estimate can be chosen equal to
$\hhatbbv(\bbu_1) := \sqrt{\lambda_1} \bbu_1$.

Besides this eigenvector approach, \emph{randomization} offers
another way to extract an approximate R-SE vector from $\hhatbbV$,
with quantifiable approximation accuracy; see e.g.,
\cite{luospmag10}. The basic idea is to generate multiple Gaussian
distributed random vectors $\bbnu \sim \ccalC \ccalN(\bb0,
\hhatbbV)$, and pick the one with the minimum error cost corresponding to the set of inlier meters $\ccalM_i:=\{\ell|1\leq \ell\leq M,~\hhata_\ell \neq 0\}$. Note that
although any vector $\bbnu$ is feasible for \eqref{rSE_wls}, it is
still possible to decrease the minimum achievable cost by rescaling
to obtain $\hhatbbv(\bbnu) = \hhatc \bbnu$, where the optimal weight
can be chosen as the solution of the following convex problem as
\begin{align}
\hhatc &
= \arg \min_{c>0}  \sum_{\ell\in\ccalM_i} w_\ell \left(z_\ell - c^2\bbnu^\ccalH\bbH_\ell\bbnu \right)^2 \nonumber \\
& = \sqrt{\frac{\sum_{\ell\in\ccalM_i} w_\ell z_\ell\bbnu^\ccalH\bbH_\ell\bbnu }{\sum_{\ell\in\ccalM_i} w_\ell (\bbnu^\ccalH \bbH_\ell\bbnu)^2}}.
\end{align}

It will be of interest to find approximation bounds for the
SDR-based R-SE approach, or, obtain meaningful conditions under
which the relaxed solution coincides with the unrelaxed one. Both
problems constitute interesting future directions for analytical
research, while the ensuing section will demonstrate the performance
improvement possible with the proposed method using numerical tests
of practical systems.


\section{Preliminary Simulations}
\label{sec:sim}

The novel SDR-based R-SE approach is tested in this section using
the IEEE 30-bus system with 41 transmission lines \cite{PSTCA}, and
compared to existing WLS methods that are based on Gauss-Newton
iterations. The software toolbox MATPOWER \cite{matpower11tps} is
used to generate the pertinent power flow and meter measurements. In
addition, its SE function \verb"doSE" has been adapted to realize
the WLS Gauss-Newton iterations following from \cite{xu_arxiv11}.
The iterations terminate either upon convergence, or, once the
condition number of the approximate linearization exceeds $10^8$,
which flags divergence of the iterates. To solve the  SDR-based R-SE
problems, the MATLAB-based optimization package \texttt{CVX}
\cite{cvx} is used, together with the interior-point method solver
SeDuMi \cite{sedumi}.

\begin{figure}[!tb]
\begin{center}
\centerline{\epsfig{file=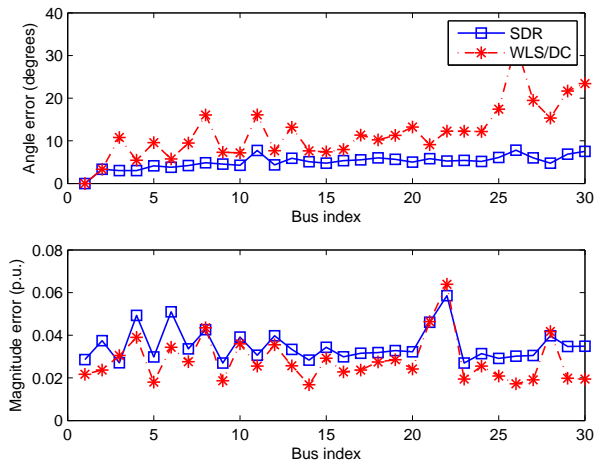,width=\linewidth}}
\caption{Comparing estimation errors in voltage magnitudes and
angles between SDR and WLS solvers at different buses.}
\label{fig:rse}
\end{center}
\end{figure}

The real and reactive power flows along all 41 lines are measured,
together with voltage magnitudes at 30 buses. AWGN corrupts all
measurements, with $\sigma_\ell$ equal to $0.02$ at power meters,
and 0.01 at voltage meters. Except for the reference bus phasor
$V_{ref}=1$, each bus has its voltage magnitude Gaussian distributed
with mean $1$ and variance $0.01$, and its voltage angle uniformly
distributed over $[-0.5\pi,~0.5\pi]$.  The empirical voltage angle
and magnitude errors per bus, averaged over 500 Monte-Carlo
realizations, are plotted in Fig. \ref{fig:rse}. In each
realization, one power flow meter measurement is randomly chosen as
a bad datum, after multiplying the meter reading by 1.2. Clearly,
the proposed algorithm greatly reduces the effects of bad data in
the estimation error in voltage phase angles (upper), which is a
more important SE performance metric than the magnitude one.

\section{Conclusions and Current Research}
\label{sec:con}

For the practical nonlinear AC power system model, uniqueness issues
and robust state estimation (R-SE) algorithms were investigated in
this paper, when outliers (bad data and/or malicious attacks) are
present. Using a sparse overcomplete outlier model, observability
and identifiability issues were quantified using the notion of
measurement distance for the quadratic measurement model. Valuable
insights and computable levels of outlier observability and
identifiability were provided for linear approximations of the
quadratically nonlinear models. A novel SDR-based scheme was also
developed by tactfully relaxing the nonconvex R-SE problem to a
convex SDP one, thus rendering it efficiently solvable via existing
interior-point methods. Preliminary numerical simulations on the
30-bus benchmark system demonstrated improved performance of the
proposed R-SE scheme.

Further enhancements to the SDR-based R-SE framework are currently
pursued toward developing more efficient and tailored solvers by
exploiting sparsity of the SDP problem structure. The measurement
distance for the power meter quadratic functions is also under
investigation using insights from nonlinear channel coding theory.


\end{document}